\pdfoutput=1
\documentclass[journal]{IEEEtran}
\usepackage{amsmath,amsthm,amssymb}
\usepackage{color}
\usepackage{graphicx}
\usepackage[caption=false,farskip=0pt,labelfont={bf}]{subfig} 
\usepackage{cite}
\usepackage{array}
\usepackage[ruled,linesnumbered]{algorithm2e}
\usepackage{multirow}
\usepackage{mathrsfs}
\usepackage{booktabs}
\usepackage[labelfont=bf]{caption}
\usepackage{balance}

\newcommand{\PreserveBackslash}[1]{\let\temp=\\#1\let\\=\temp}

\newcolumntype{C}[1]{>{\PreserveBackslash\centering}m{#1}}
\newcolumntype{R}[1]{>{\PreserveBackslash\raggedleft}m{#1}}
\newcolumntype{L}[1]{>{\PreserveBackslash\raggedright}m{#1}}

\begin{document}
\title{An Efficient Early-breaking Estimation and Tree-splitting Missing RFID Tag Identification Protocol}

\author{Lijuan Zhang, Mingqiu Fan, Chunni Yu, Lei Lei
	\thanks{The authors are with the College of Electronic and Information Engineering, Nanjing University of Aeronautics and Astronautics, Nanjing 211106, China (e-mail: lijuanzhang6@gmail.com; fanmingqiu@foxmail.com;  yuchunni0228@163.com;leilei@nuaa.edu.cn )
}}

\maketitle

\begin{abstract}
Recent statistics have demonstrated that missing items have become the main cause of loss for retailers in inventory management. To quickly identify missing tags, traditional protocols adopt Aloha-based strategies which take a long time, especially when the number of tags is large. Among them, few works considered the effect of unexpected unknown tags on the missing tag identification process. With the presence of unknown tags, some missing tags may be falsely identified as present. Thus, the system's reliability is hardly guaranteed. In this work, we propose an efficient early-breaking estimation and tree-splitting-based missing tag identification (ETMTI) protocol for large-scale RFID systems. In ETMTI, a new early-breaking estimation and deactivation method is developed to effectively estimate the number of unknown tags and deactivate them within a short time. Next, a new tree-splitting-based missing tag identification method is proposed to quickly identify missing tags with a B-ary
splitting tree. Besides, a bit-tracking response strategy is designed to further reduce the time cost. The optimal parameters, time
cost, and false negative rate of ETMTI are analyzed theoretically. Simulation results are presented to demonstrate that the proposed ETMTI protocol takes a smaller time and has a lower false negative rate than the best-performing benchmarks.
\end{abstract}


\section{Introduction}
\IEEEPARstart{R}{ecently}, radio frequency identification (RFID) has been widely applied in many domains, such as logistics, manufacturing, pharmaceutical industry, and so on \cite{Wang19_bac,Chen17_bac}.  As one of the key perception technologies that enable Internet of Things (IoT) networks, RFID exhibits many advantages, including non-contact, non-visual reading, strong anti-interference ability, high reliability, and capablility of working in harsh environments \textit{etc}. 
According to a study conducted by the National Retail Federation \cite{nrf.org}, retailers suffered \$94.5 billion in 2021 due to shoplifting, inventory loss, internal theft, management errors,
supplier fraud, and other reasons. Missing items have become the main cause of loss for retailers in inventory management. In these applications, readers are used to monitor tags in stock frequently for goods management and inventory. 

To effectively identify the missing items, many missing tag identification protocols, including probabilistic and deterministic ones, are proposed. On the one hand, probabilistic protocols implement lightweight operations to detect the missing tag event with predefined reliability \cite{Shahzad16_RUN, Yu17_BMTD, Zhang20_CBMTD}. These works usually take a short time to discover a missing tag event, but they cannot provide ID information of the missing tags. On the other hand, deterministic protocols give ID information of missing tags \cite{Li13_THP, Liu14_MMTI,Liu15_SFMTI,Zhang17_PCMTI,Su23_CRMTI}.  Making use of the hash mapping method, these protocols assign known tags to different slots and identify missing tags by checking whether there is a tag response in the expected singleton slot. If no response is detected, the corresponding tag is missing. Otherwise, it is a present one. To further improve the identification efficiency, some recent works considered to use bit-tracking technology, such as the pair-wise collision-resolving missing tag identification (PCMTI) protocol \cite{Zhang17_PCMTI} and the collision resolving-based missing tag identification (CRMTI) protocol \cite{Su23_CRMTI}.
However, these works assumed that all tags within range are known to the reader without considering any unexpected unknown tags. 

In practical scenarios, some unknown tags may present  and  affect the identification of missing tags. With the presence of  unknown tags, a missing tag may be misidentified as a present one if the unknown tag is assigned to the expected singleton slot and replies a 1-bit short message to the reader. In the literature, a few works considered the effect of unknown tags and tried to deactivate them, such as the two-phased bloom filter-based missing tag detection (BMTD) protocol \cite{Yu17_BMTD} and the efficient and reliable missing tag identification (ERMI) protocol \cite{Chen17_ERMI}. 
In general, existing missing tag identification protocols have the following limitations:
\begin{itemize}
	\item Since the reader has no prior information about unknown tags, an efficient unknown tag number estimation method is of great importance to guarantee the required reliability. For time-saving consideration, existing works either lack the estimation process  or only provide a rough estimation that the required reliability is not always guaranteed;
	\item Existing works implement Aloha-based strategies to identify missing tags. In each frame, unidentified tags are randomly assigned to slots with hash mapping. None of them considered making use of information in the preceding frames. The slot information is not fully used and the time efficiency needs further improvement; 
	\item In previous works, tag replies to the reader with a one-bit short response. To reduce the time cost, several works considered using customized responses with the help of bit-tracking technology. However, there still exist many short response slots that lower the time efficiency.  
\end{itemize}

In this work, an efficient early-breaking estimation and tree-splitting-based missing tag identification (ETMTI) protocol is proposed for large-scale RFID systems. In ETMTI, two new methods are developed to enhance the unknown tag deactivation and missing tag identification process, respectively. The major contributions of this work are in four folds as in the following.
\begin{itemize}
	\item [1)] A new early-breaking estimation-based unknown tag deactivation (EBUD) method is developed to  estimate the number of unknown tags and deactivate them within a short time. The early-breaking factor is chosen to balance time cost and estimation accuracy, and the number of frames is determined to guarantee the required reliability;
	\item[2)] A new  tree-splitting-based missing tag identification (TSMTI) method is designed to effectively identify missing tags. In TSMTI, the $B$-ary splitting tree method is developed to accelerate the identification process.  The optimal frame factor and branch number in TSMTI are derived theoretically to minimize the execution time;
	\item[3)] A bit-tracking response strategy that allows simultaneous replies of multiple tags is developed to accelerate the identification process. With customized tag	responses,  the reader can identify multiple tags in one slot,	which greatly reduces identification time.
	\item[4)] Theoretical analysis is conducted to optimize the parameter settings and derive the expressions of time cost in each phase. Numerous simulation results are presented to demonstrate the effectiveness of ETMTI. Compared with existing benchmark works, ETMTI takes a shorter identification time and a lower false negative rate to identify missing tags.    
\end{itemize}

The remainder of this work is organized as follows: Section \ref{sec_related} reviews the most related works on missing tag identification.  Section \ref{sec_preliminary} gives the system model of this work. 
In Section \ref{ours}, the proposed ETMTI protocol is described in detail. Then, theoretical analysis is conducted in Section \ref{sec_ana}. Simulation results are presented in Section \ref{sec_eva}. Finally, some concluding remarks are made in Section \ref{sec_conclusion}.

\section{Related works} \label{sec_related}

In this section, we first introduce the traditional missing tag identification protocols with only known tags. Next, the related works that deal with unknown tags are reviewed. 

\subsection{missing tag identification with only known tags}

In the last decade, many missing tag identification protocols are proposed to specify the ID information of missing tags from the known ones. Li \textit{et al.} first proposed the two-phased protocol (TPP) and two-hush protocol (THP) \cite{Li13_THP}. Next, Liu \textit{et al.} proposed a multi-hashing-based missing tag identification (MMTI) protocol to improve the utilization of each frame \cite{Liu14_MMTI}. With multiple hash assignments in MMTI, many expected empty or collision slots are changed into expected singleton slots so that more tags can be identified in a frame.  Making use of multiple hash seeds, the slot-filter-based missing tag identification (SFMTI) protocol \cite{Liu15_SFMTI} reconciles expected collision slots with 2 or 3 tags into singleton slots to further improve the utilization of a frame. Later on, some similar protocols that make use of the reconcilable collision slots are proposed, such as the coarse-grained inventory list-based stocktaking protocol \cite{Zhu20_CLS} and the collision reconciliation and data compression algorithm \cite{Wang21_GP}. 

Considering the requirements of practical applications, Chen \textit{et al.} proposed an improved vector-based missing key tag identification (iVEKI) protocol \cite{Chen18_VEKI} to deactivate ordinary tags and identify missing key tags, separately. Thus, the missing more valuable key tags can be identified more efficiently. Considering privacy-leakage prevention, Wang \textit{et al.} made use of the group-based and collision-reconciled protocols to identify missing tags in blocker-enabled systems \cite{Wang19_blocker}. In \cite{Yu20_P2P}, Yu \textit{et al.} proposed the point-to-multipoint (P2M) and collision-free point-to-point (P2P) protocols to reduce communication cost. However, most of these works concentrate on improving  frame utilization with the help of either multiple hash assignments or collision reconciliation strategies. Much useful information is wasted.

In recent research, a few missing tag identification protocols are considered to use bit-tracking technology. With Manchester encoding, the reader is capable of detecting the positions of colliding bits in the received collision message and retrieving useful information in the collision slot. Actually, bit-tracking has been widely applied in many tag anti-collision protocols, such as  the $M$-ary collision tree protocol \cite{Zhang18_MCT}, efficient bit-detecting protocol \cite{Zhang18_EBD}, modified dual prefixes matching mechanism \cite{Su20_MDPM} and so on. For missing tags, PCMTI verifies the presence of two tags in each slot with the help of bit-tracking \cite{Zhang17_PCMTI}. To further improve the identification efficiency, CRMTI takes advantage of both bit-tracking and collision resolving technologies to allow customized tag responses in the reconcilable collision slots\cite{Su23_CRMTI}. These strategies can reduce time costs to some extent, but they did not make full use of the bit-tracking technology. 

\subsection{missing tag identification with unknown tags}

Many works assumed that reader knows the ID information of all present tags within the reading range, which is unrealistic in most applications. In the literature, Shahzad \textit{et al.} took the first step to consider the effect of unknown tags and proposed two RFID monitoring protocols with unexpected tags (RUN) \cite{Shahzad16_RUN}, i.e., RUN$_\text{D}$ and RUN$_\text{I}$ for probabilistic and deterministic missing tag identifications, respectively. In their work, multiple frames with different seeds are executed to reduce the effect of unknown tags, and the number of unknown tags is estimated from the executed frames and used to optimize the frame parameters to reduce time cost. Although RUN did not take any additional frames for estimation, the execution of all slots in each frame takes a long time. In \cite{Xie17_FCS}, Xie \textit{et al.} proposed a fast continuous scanning (FCS) protocol that uses multiple categories filter to detect unknown tags and skip the non-singleton slots to improve the identification efficiency.

To further reduce the effect of unknown tags, Chen \textit{et al.} proposed two ERMI protocols \cite{Chen17_ERMI} and separated the process into unknown tag deactivation and missing tag identification phases.
In the first phase, reader estimates the number of unknown tags and deactivates them. With the estimated tag number and predefined reliability, the frame parameters are optimized to minimize the execution time. In the second phase, the traditional hash assignment method is used for missing tag identification. However, the required reliability of ERMI is not always guaranteed, especially when the number of unknown tags is large. Similarly, Yu \textit{et al.} introduced the BMTD protocol to deactivate unexpected unknown tags and then to detect tag missing events \cite{Yu17_BMTD}. Following up, a compressed filter-based BMTD (CBMTD) protocol is proposed to further reduce the time cost \cite{Zhang20_CBMTD}. Wang \textit{et al.} also proposed a near-optimal protocol (OPT-G) \cite{Wang21_OPT} to notify the group ID of known tags in the presence of unexpected unknown tags.

Recently, some unknown tag number estimation protocols are proposed. In \cite{Xiao17_Churn}, Xiao \textit{et al.} studied the churn estimation problem in dynamic RFID systems and proposed three churn estimators to estimate the numbers of missing, present, and unknown tags, separately. They used the state changes caused by missing and unknown tags to estimate the number of dynamic tags, but the slots with both missing and unknown tags are wasted. In \cite{Liu18_SEBU}, Liu \textit{et al.} proposed a simultaneous estimation of the blocked tag size and the unknown tag size (SEBU) protocol to facilitate the identification of blocked RFID tags.  Xi \textit{et al.} implemented single-slot count (SCT) and time slot reuse (TSR) strategies in SSR (SCT+TSR) protocol to estimate the numbers of missing and unknown tags simultaneously \cite{Xi21_SCT}. 
Considering unreliable channels, Wang \textit{et al.} proposed a cardinality estimation scheme (CEUT) to estimate the number of unknown tags in the presence of known tags\cite{Wang22_CEUT}. However, these works focus on increasing the estimation accuracy of unknown tag numbers and  the time cost is high.  

Moreover, some special strategies are introduced to mitigate the effect of unknown tags. In \cite{Wang20_OMTI}, Wang \textit{et al.} proposed  an order-based missing tag identification (OMTI) protocol  to dynamically assign each tag an exclusive slot. With offline serialization and online identification, the effect of unknown tags is reduced. In \cite{Chen22_HDMI}, Chen \textit{et al.} presented an efficient and accurate protocol to identify missing tags in high dynamic RFID systems. They combined the reply slot location and reply bits of tags for simultaneous missing tag identification and unknown tag filtering. Besides, some unknown tag identification protocols are proposed to separate known and unknown tags \cite{Chu22_EUTI,Liu20_CSD,Liu20_det}.
In general, existing works take some strategies to reduce the effect of unknown tags. Whereas they usually take the basic hash assignment method to identify missing tags which takes a long time to meet the high-reliability requirement.

\section{System model}\label{sec_preliminary}

\begin{figure}[ht]
	\centering
	\includegraphics[width=0.5\textwidth]{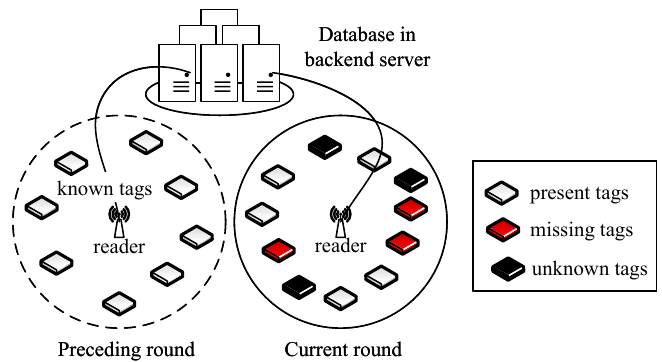}
	\caption{System model of a large-scale RFID system with both known and unknown tags. Note that the ID information of known tags is stored in the backend database, and the reader has no prior information about unknown tags.}\label{fig_model}
\end{figure}

 
This work considers a typical large-scale RFID system with a reader, a backend server, and numerous tags as in Fig. \ref{fig_model}. Tags are attached to objects for ease of identification, classification, sorting, and other inventory management. For simplicity, each object is assumed to have one tag and is represented by the corresponding tag ID.  The reader is in charge of monitoring all tags within its reading range and uploads the collected ID information to the database in the backend server. Reader can also retrieve information of tags stored in the database via a high-speed channel.  The backend server has powerful communication, computation, and storage capabilities that can effectively assist reader to monitor tags. 
Each tag has a unique ID and is capable of simple computation operations as in \cite{Chen17_ERMI,Su23_CRMTI}, such as random number generation, lightweight hash function, modulus operation, and so on. 

With stock management, the ID and other information of new tags are collected and recorded in the backend database with traditional tag anti-collision protocols in warehouse entry. In the system, the set of tags may dynamically change because of management faults or theft. For example, some tags may be taken to the wrong zone and newly appear in reader's reading range; some may be stolen or mistakenly move out of the reading range. Therefore, reader has to frequently monitor all tags within range to identify missing ones as soon as possible.  

To efficiently identify missing tags, reader verifies the state of each tag by comparing the collected tag response with the backend database. Since the reader can retrieve all tags' ID information from the backend database, we denote the tags stored in the database by \textit{known tags}. A \textit{reading round} is referred to as the process in which reader verifies states of all known tags. As is shown in Fig.\ref{fig_model}, if a known tag is still within the reading range in the current round, the tag is referred to as \textit{present tag}; otherwise, it is a \textit{missing tag}. Besides, if a tag newly appears in the reading range, \textit{i.e.}, there is no information in the database, it is an \textit{unknown tag}. 


Denote the numbers of known and unknown tags by $\mathcal{K}$ and $\mathcal{U}$, respectively. The number of  missing tags is represented by $\mathcal{M}$. Affected by the presence of unexpected unknown tags, a missing tag may be falsely identified as present. Let $\mathcal{M}_{fls}$ indicate the number of falsely identified missing tags. \textbf{We define the false negative rate $\mathcal{r}_{fn}$ be the number of falsely identified missing tags to the total number of missing tags.} Given a required reliability $\alpha$, reader has to identify all missing tags in $\mathcal{M}$ and the following inequality should be guaranteed, \textit{i.e.},
\begin{eqnarray}\label{eq_fn}
\mathcal{r}_{fn}=\frac{\mathcal{M}_{fls}}{\mathcal{M}}< 1-\alpha.
\end{eqnarray}

The \textbf{\textit{main object}} of this work is to reduce time cost and false negative rate in missing tag identification with the presence of unknown tags in large-scale RFID systems.

\section{Proposed ETMTI protocol} \label{ours}

\begin{figure}[htp]
	\centering
	\includegraphics[width=0.45\textwidth]{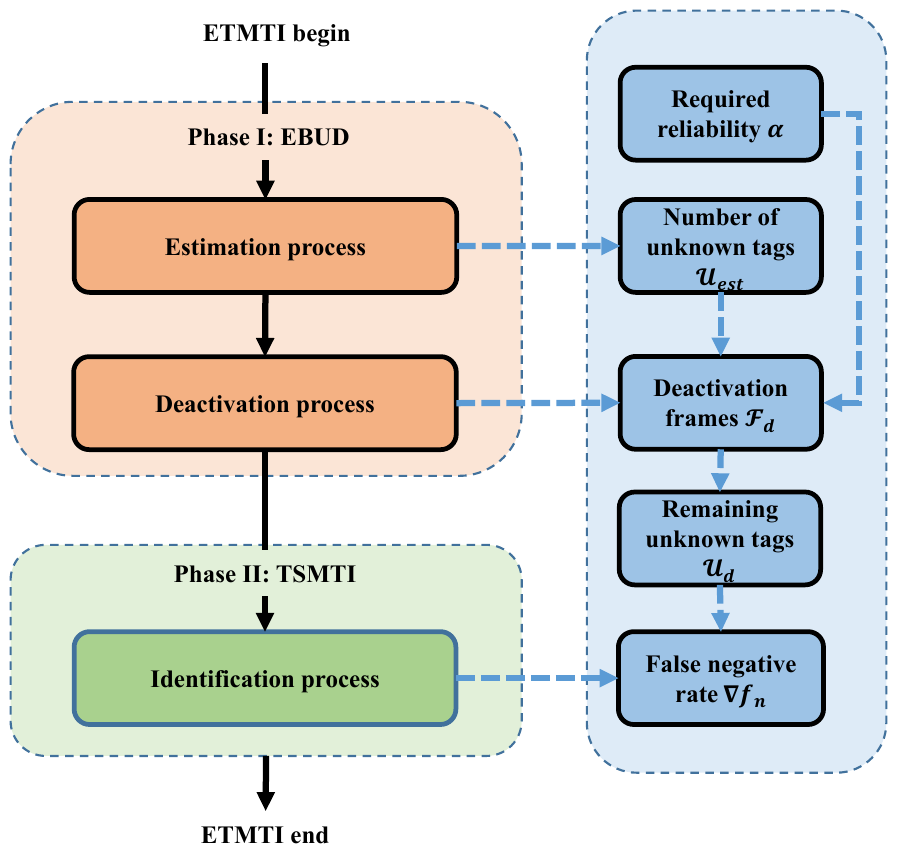}
	\caption{Schematic of ETMTI: (1) in Phase I, reader estimates number of unknown tags and deactivate them. Number of frames needed in Phase I is determined to meet the required number of remaining unknown tags that participate in Phase II;  (2) in Phase II, reader identifies missing tags with tree-splitting method. Number of remaining unknown tags allowed to participate in this phase is dereclty determined on number of deactivation frames and indirectly determined based on the estimation and required reliability.\label{fig_para}} 					
\end{figure}

In this section, we describe the proposed ETMTI protocol in detail. The identification process of ETMTI consists of two phases, \textit{i.e.}, unknown tag deactivation, and missing tag identification phases. As is illustrated in Fig. \ref{fig_para}, a new early-breaking estimation-based unknown tag deactivation (EBUD) method is developed in Phase I to effectively estimate the number of unknown tags and deactivate them. With EBUD, most unknown tags can be deactivated in a very short time. In Phase II, a new tree-splitting-based missing tag identification (TSMTI) method is developed to effectively identify missing tags and deactivate the remaining unknown ones. With tree-splitting, the identification time is greatly reduced and the reliability is further improved.

\subsection{Phase I: early-breaking estimation-based unknown tag deactivation} \label{sec_EBUD}
 
\begin{figure}[htp]
	\centering	
	\includegraphics[width=0.48\textwidth]{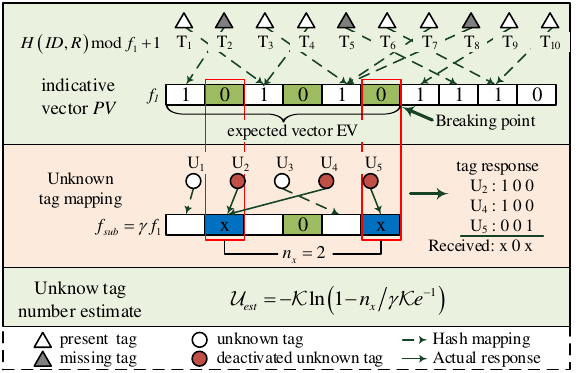}
	\caption{Early-breaking estimation-based unknown tag deactivation.\label{fig_alg_est}  }			
\end{figure}

In this phase, reader executes a new EBUD algorithm to estimate the number of unknown tags in the first frame and deactivate them in subsequent frames. 
In the $i$-th frame of this phase, the reader first assigns known tags with hash mapping to construct an indicative vector $PV$. In detail, it generates the random seed $R$, sets frame size $f_i=\mathcal{K}$ and calculates slot index for tag $T_j$ by
\begin{eqnarray}\label{eq_hash}
	s~=~H(ID_j,~R)~mod~f_i~+~1,
\end{eqnarray}
where $H()$ is a hash function. Then, it generates $PV$ with $f_i$ bits zeros and sets the $s$-th bit to be ``1", representing that the $s$-th slot is an \textit{expected non-empty slot}. If there is no tag assigned, the reader sets the corresponding bit to be ``0", denoting an \textit{expected empty slot}. As is shown on top of Fig. \ref{fig_alg_est}, the constructed $PV=$``1 0 1 0 1 0 1 1 1 0",  \textit{i.e.}, only the 2nd, 4-th, 6-th and 10-th slots are expected empty slots. 

To effectively estimate the number of unknown tags, a new early-breaking estimation method is introduced.  As is illustrated in Fig. \ref{fig_alg_est}, reader sets the breaking point to break $PV$ into two parts, and the first $f_{sub}$ bits are expressed as the expected vector $EV$.  Note that $f_{sub}=\lceil \gamma f_1\rceil$, where $\gamma$ is the early-breaking factor ranging in $[0,1]$ and $\lceil \cdot \rceil$ is the ceiling function.  Then, it broadcasts $Querye(R,~f_1,~EV)$ command to inform tags with the random seed, frame size, and the expected vector.  It should be noted that transmitting only the subvector of $PV$ reduces time cost. 

Since all known tags will be assigned to the ``1" bit positions, they will keep silent and wait for the next query command. Only unknown tags might map to the ``0" bit positions, hence the reader can estimate the number of unknown tags through checking response information of the expected empty slots.   
After receiving the \textit{Querye} command, a tag calculates the slot index $s$ with \eqref{eq_hash} and checks the corresponding bit in $EV$.  If $EV(s)$ is ``1" or $s$ is greater than the length of $f_{sub}$, it will keep silent in the current frame.  If $EV(s)$ is ``0", the tag confirms that it is an unknown tag and constructs its response string with bit-tracking response method.   Denote the number of ``0"s in $EV$ by $\overline{n}_0$ and the number of ``0"s prior to the $s$-th position by ${n_0}$.  The tag first generates a $\overline{n}_0$ bit response string $R_{str}$ by setting the $(n_0+1)$-th bit to ``1" and other bits be ``0"s.  Then, it replies $R_{str}$ to the reader and deactivates itself immediately.  
More specifically, as is shown in the middle of Fig. \ref{fig_alg_est}, tags $U_1$ and $U_3$ are assigned into ``1" bit positions of $EV$ that they  will keep silent in the current frame. Since tags $U_2$, $U_4$ and $U_5$ are assigned into the ``0" bit positions, they will reply and deactivate themselves in this frame.  Taking $U_5$ as an example, it constructs the response string as ``001", since there are 3 ``0"s in $EV$ and tag $U_5$ is assigned in the $3$-rd ``0" bit position.  Similarly, the response strings of tags $U_2$ and $U_4$ are the same, \textit{i.e.}, ``100". With bit-tracking technology, the received message at the reader side in this frame is ``x0x", where ``x" refers to a colliding bit. Notice that if there is only one tag response, the received message will have one ``1" bit which is also regarded as an ``x". 

By calculating the number of ``x"s $n_x$, reader estimates  the number of unknown tags.  Since tags are randomly assigned into slots, the probability that a tag is assigned to a specific slot is $1/f_1$. If the reader detects an ``x" in the received message, it knows that at least one unknown tag replies in the \textit{expected empty slot}.  Recalling the construction of $PV$, the probability that no known tags are assigned to a specific bit position in $PV$ is expressed as
\begin{eqnarray} \label{eq_p0}
p_0=\bigg(1-\frac{1}{f_1}\bigg)^{\mathcal{K}}\approx e^{-\frac{\mathcal{K}}{f_1}}\approx e^{-1}.  
\end{eqnarray}
 Similarly, the probability that at least one unknown tag are assigned in a specific position is calculated as
\begin{eqnarray}\label{eq_px0}
 p_{u}=\bigg[1-\bigg(1-\frac{1}{f_1}\bigg)^{\mathcal{U}}\bigg]\approx1-e^{-\frac{\mathcal{U}}{f_1}}=1-e^{-\frac{\mathcal{U}}{\mathcal{K}}}. 
\end{eqnarray}
Then, the probability that reader detects an ``x" is given by 
\begin{eqnarray}\label{eq_px}
p_{\text{x}}&=&p_0\cdot p_{u}\approx e^{-1}\bigg(1-e^{-\frac{\mathcal{U}}{f_1}}\bigg)=e^{-1}\bigg(1-e^{-\frac{\mathcal{U}}{\mathcal{K}}}\bigg).
\end{eqnarray}
The expectation of the number of ``x"s in the received message is calculated by $E(n_x)=\gamma f_1 \cdot p_{\text{x}}$. 
Suppose the actual number of ``x"s $n_x$ is approximately $E(n_x)$, $n_x\approx E(n_x)=\gamma f_1 \cdot p_{\text{x}}$. Substituting it to (\ref{eq_px}), the estimated number of unknown tags is calculated 
\begin{eqnarray}\label{eq_est}
\mathcal{U}_{est}=-\mathcal{K}\ln \bigg(1-\frac{n_x}{\gamma \mathcal{K}\cdot e^{-1}}\bigg).
\end{eqnarray}

In subsequent frames, the reader does similar operations to deactivate unknown tags. It assigns known tags into slots to construct the indicative vector $PV$ and broadcasts $Queryd(R,~f_i,~PV)$ command to tags.  On receiving this command, tags that are assigned into the ``0" bit positions in $PV$ do hash mapping operations and deactivate themselves immediately. The deactivated unknown tags will not participate in Phase II.  In general, the main discrepancies between the estimation and deactivation processes are in two folds. Firstly, in $Querye$ command, an expected vector $EV$ copied from the first $\gamma f_1$ bits from $PV$ is transmitted. Whereas in $Queryd$ the full $PV$ string is transmitted. Secondly, after receiving $Querye$ command, tags that are assigned into ``0" bit positions in $EV$ will reply to the reader. However, after receiving $Queryd$ commands, tags will not reply to reader, \textit{i.e.}, there are only reader's commands transmitted in each frame in the deactivation process. Tags that are assigned to the expected empty slots will deactivate themselves and keep silent.

 \begin{figure*}[htp]
 	\centering
 	\includegraphics[width=1\textwidth]{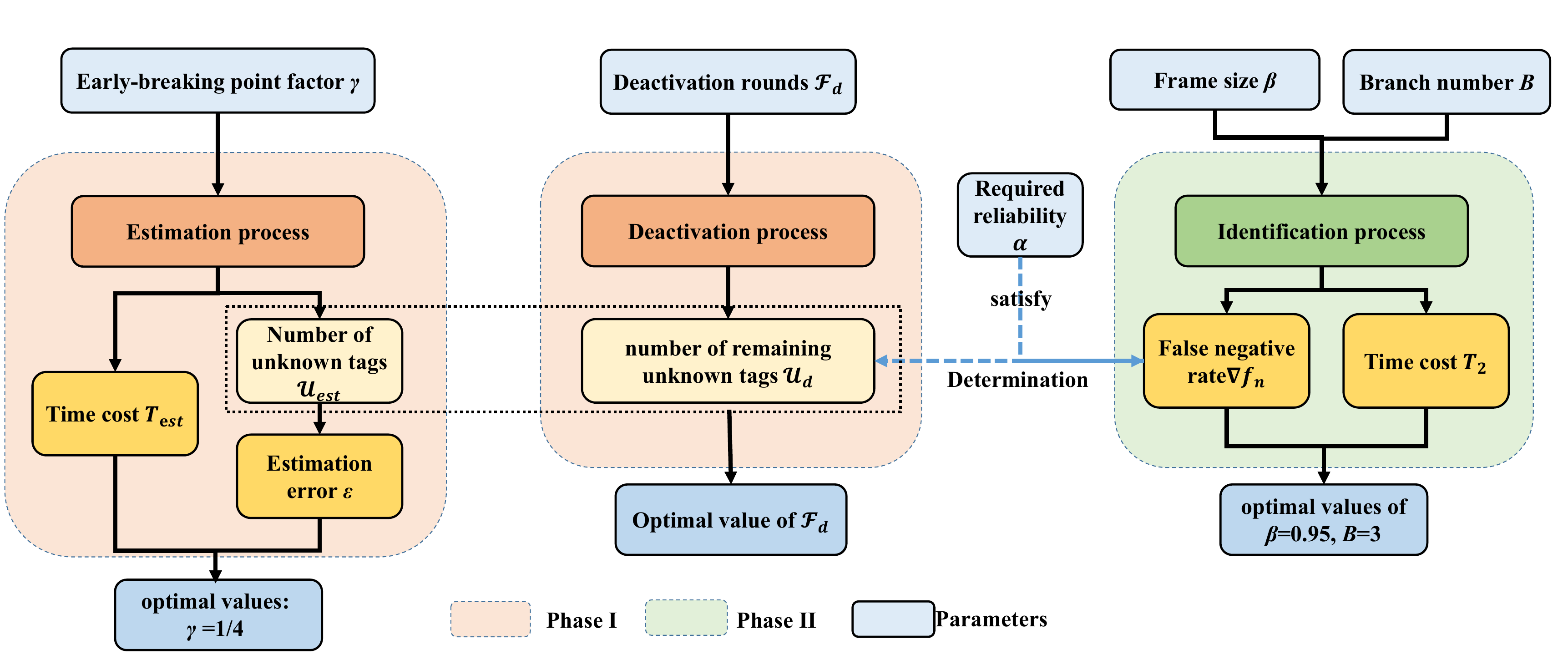}
 	\caption{The logic diagram of perfomance analysis: firstly, time cost of Phase I is analyzed and the early-breaking factor $\gamma$ is determined through balancing estimation error $\epsilon$ and time cost $T_{est}$; secondly, time cost $T_2$ and false negative rate $\mathcal{r}_{fn}$ of Phase II is analyzed and the optimal frame parameter $\beta$ and branch number $B$ are obtained; finally, the number of frames in Phase I $\mathcal{F}_d$ is determined to deactivate enough unknown tags based on the estimated unknown tag number $\mathcal{U}_{est}$ and required reliability $\alpha$.\label{fig_paraall}} 					
 \end{figure*} 

\subsection{Phase II: tree-splitting-based missing tag identification} \label{sec_TSMTI}

\begin{figure}[htp]
	\centering	
	\includegraphics[width=0.48\textwidth]{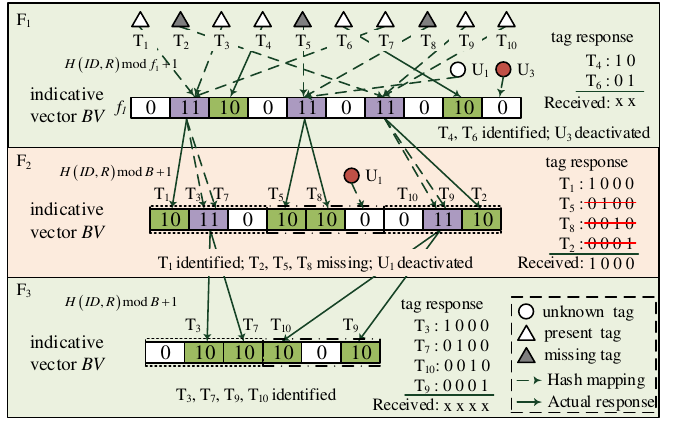}
	\caption{$B$-ary tree-splitting-based missing tag identification.\label{fig_alg_iden}  }			
\end{figure}

In this phase, the reader executes the $B$-ary tree-splitting method to quickly identify missing tags and deactivate the remaining unknown tags.  What's more, the first frame is also different from the subsequent frames. 
In the first frame, the reader generates a random hash seed, sets frame length, and assigns known tags with hash mapping to construct the indicative vector $BV$ as in Fig. \ref{fig_alg_iden}. Different from Phase I, three states should be indicated in $BV$: (i) If there is no tag assigned in a specific segment, this is an \textit{expected empty slot} and  denoted by a single ``0" bit; (ii) If only one tag is assigned, this is an \textit{expected singleton slot} and represented by ``10"; (iii) Otherwise, this is an \textit{expected collision slot} and denoted by ``11". For example, in $F_1$ of Fig. \ref{fig_alg_iden}, the $3\text{-rd~and}~9\text{-th}$ slots are two expected singleton slots, the $2\text{-nd},~5\text{-th},~\text{and}~7\text{-th}$ slots are three expected collision slots, and others are expected empty slots. Then the constructed indicative vector  $BV=$``0 11 10 0 11 0 11 0 10 0". 

To facilitate the tree-splitting process, the reader keeps a counter for each known tag, \textit{i.e.}, $Ac(T_j)$ for tag $T_j$. If tag $T_j$ is assigned into an \textit{expected singleton slot}, the reader sets $Ac(T_j)=0$; if it is assigned into an \textit{expected collision slot}, the reader calculates the number of ``11" segments prior to the assigned position (denoted by $\mathcal{X}_{11}$) , and sets $Ac(T_j)=\mathcal{X}_{11}+1$. Then, the reader broadcasts $Querym(R,~f_1,~BV)$ and waits for tag responses. After receiving this command, tag $T_j$ does the same hash mapping operations as the reader and checks the corresponding segment in $BV$ as follows:
\begin{itemize}
	\item If the assigned segment is ``10", the tag sets $Ac(T_j)=0$ and prepares an $\overline{\mathcal{X}}_{10}$ bits response string $R_{str}$, where $\overline{\mathcal{X}}_{10}$ is the number of ``10"s in $BV$. For instance, in frame $F_1$ of Fig. \ref{fig_alg_iden}, $Ac(T_4)=Ac(T_6)=0$, and $\overline{\mathcal{X}}_{10}=2$; 
	\item If the assigned segment is ``11", the tag does similar operations as the reader to obtain $\mathcal{X}_{11}$, and sets $Ac(T_j)=\mathcal{X}_{11}+1$. As is shown in frame $F_1$ of Fig. \ref{fig_alg_iden},	$Ac(T_1)=Ac(T_3)=Ac(T_7)=1$, $Ac(T_5)=Ac(T_8)=Ac(U_1)=2$ and $Ac(T_2)=Ac(T_9)=Ac(T_{10})=3$;
	\item If the assigned segment is ``0", the tag determines that it is an unknown tag and will be deactivated. In frame $F_1$ of Fig. \ref{fig_alg_iden}, we can observe that $U_3$ is deactivated.
\end{itemize}

For a tag with $Ac(T_j)=0$, it sets all bits of $R_{str}$ to be zero. It then counts the number of ``10"s prior to its assigned segment in $BV$, and sets the corresponding bit in $R_{str}$ to be bit ``1".  For example, in $F_1$ of Fig. \ref{fig_alg_iden}, tag $T_4$ is assigned into the first ``10" segment in $BV$. It sets $R_{str}=$``10". Similarly, tag $T_6$ is assigned into the second ``10" segment in $PV$, so that it sets  $R_{str}=$``01". Then, the two tags reply $R_{str}$ and keep silent. After receiving tag responses, the reader decodes the received message as ``xx" and confirms that tags $T_4$ and $T_6$ are present tags.

In subsequent frames, the reader identifies missing tags with a $B$-ary tree. In detail, the reader divides the $i$-th frame ($i\geq$2) into multiple groups based on the number of expected collision slots in the ($i$-1)-th frame. Each group consists of $B$ slots. The group index of each tag is determined by its counter value $Ac$. In each group, the reader assigns tags with $s=H(ID,~R)~mod~B~+1$, and constructs indicative vector $BV$ by concatenating the slot states in all groups. It then updates the counter values of all tags based on the constructed $BV$. For example in $F_2$ of Fig. \ref{fig_alg_iden}, with $B$=3, the reader assigns $T_1$, $T_3$ and $T_7$ in the first three slots because their $Ac$=1; $T_5$ and $T_8$ with their $Ac$=2 are assigned into the second group; $T_2$, $T_9$ and $T_{10}$ are assigned into the third group. The constructed indicative vector $BV$=``10 11 0 10 10 0 0 11 10". Since $T_3$ and $T_7$ are assigned in the first expected collision
slot; $T_9$ and $T_{10}$ are assigned in the second expected collision slot; other tags are assigned in the expected singleton slots, tags update their counter values as 
$Ac(T_1)=Ac(T_5)=Ac(T_8)=Ac(T_2)=0$, $Ac(T_3)=Ac(T_7)=1$, $Ac(T_9)=Ac(T_{10})=2$.  Next, the reader broadcasts $Querym(R,~B,~BV)$ to tags.

On receiving reader's $Querym$ command, tag $T_j$ does similar hash operations as the reader and checks corresponding segments of the  $Ac(T_j)$-th group in $BV$. Then it operates similarly to the tags in the first frame. If the tag is assigned into an expected singleton slot, it first checks the number of ``10"s in $BV$, denoting by $\overline{\mathcal{X}}_{10}$ and generates a response string $R_{str}$ with $\overline{\mathcal{X}}_{10}$  zero bits. It then checks the number ``10"s prior to its assigned position and sets the corresponding bits in $R_{str}$ into ``1" and reply to the reader. If the tag is assigned into an expected collision slot. It calculates the number of ``11"s prior to its assigned position and updates $Ac$ accordingly. If the tag is assigned to an expected empty slot, it deactivates itself. 

For example, in $F_2$ of Fig. \ref{fig_alg_iden}, tags $T_1,~T_3,~\text{and}  ~T_7$ do hash mapping operations and check the first three segments ,\textit{i.e.}, the first group, in $BV$. Tag $T_1$ is assigned to an expected singleton slot and tags $T_3$ and $T_7$ are assigned into an expected collision slot. Tag $T_1$ check the number of ``10"s, generates $R_{str}=``1000"$ and replies to the reader. Tags $T_3$ and $T_7$ check the number of ``11"s prior to their assigned position and update their counter values as  $Ac(T_3)=Ac(T_7)=1$. In the $4\text{-th}$ to $6\text{-th}$ segments, tag $U_1$ is assigned into an expected empty slot that will be deactivated. In the $7\text{-th}$ to $9\text{-th}$ segments, tags $T_9$ and $T_{10}$ are assigned into the same expected collision slot. They update  $Ac(T_9)=Ac(T_{10})=2$. Since tags $T_2$, $T_5$ and $T_8$ are missing, only tag $T_1$ will reply in this frame. 

After receiving tags' responses, the reader determines that tag $T_1$ is present, and tags $T_2$, $T_5$, and $T_8$ are missing. Similarly, the reader confirms that tags $T_3$, $T_7$, $T_9$ and $T_{10}$ are present  in $F_3$. If there are no collision slots in $F_3$, it means all tags are identified. Then the reader terminates the current reading round. Otherwise, it splits collision slots and repeats the identification process in subsequent frames.
With tree-splitting, colliding tags are more easily separated and the identification process is effectively accelerated. 

\section{Performance analysis}\label{sec_ana}

In this section, we first analyze the deactivation phase and optimize the early-breaking factor $\gamma$ to balance the time cost and estimation error of EBUD. Next, we analyze the identification phase and optimize the frame parameter $\beta$  and the branch number $B$.  Then, the false negative rate of the identification phase is analyzed. Since the false negative rate is affected by the number of unknown tags participating in Phase II, number of frames needed in Phase I is determined by making use of the estimated unknown tag number and the required reliability to deactivate enough unknown tags.    More specifically, Fig. \ref{fig_paraall} illustrates the main logic of our analysis. 

\subsection{Time cost of Phase I} \label{sec_paraIe}
In Phase I, a new EBUD method is developed to effectively estimate the number of unknown tags and deactivate them.  Time cost of EBUD is given by
\begin{eqnarray}\label{eq_T1}
	T_1=T_{est}+T_{dea},
\end{eqnarray}
where $T_{est}$ and $T_{dea}$ are the time costs of the estimation and deactivation processes, respectively. 

In the estimation process, each frame consists of the transmission of reader's $Querye()$ command and unknown tags' responses. As given in Section \ref{sec_EBUD}, $Querye(~R,~f_1,~EV)$ command consists of a 4-bit command type string, a 16-bit hash seed, a 16-bit frame size, and a $\gamma f_1$-bit expected vector.  On the tag side, unknown tags are assigned to expected empty slots that will reply immediately.  With bit-tracking response, the length of a response message is the number of expected empty slots indicated in $EV$. Since the probability of a specific slot to be empty is $(1-\frac{1}{f_1})^\mathcal{K}$, the number of expected empty slots is $\gamma f_1(1-\frac{1}{f_1})^\mathcal{K}$.  Thus, the time cost is given by
\begin{eqnarray}\label{eq_test}
T_{est}=&\bigg\{\!\underbrace{ \bigg\lceil\frac{\gamma f_1+16\cdot 3+4}{96} \bigg\rceil}_{reader~request} 
&+\underbrace{ \bigg\lceil\frac{\gamma f_1\big(1-\frac{1}{f_1}\big)^{\mathcal{K}}}{96} \bigg\rceil}_{tag~responses}\! \bigg\}t_{id},
\end{eqnarray} 
where $f_1=\mathcal{K}$ and $t_{id}$ is time cost for transmitting a 96-bit string. It should be noted that both reader's request command and tags' responses are divided into 96-bit segments to facilitate transmission.   

In the estimation process, two indexes, $T_{est}$ and estimation error $\epsilon$ are adopted to determine the early-breaking factor $\gamma$.   Define estimation error as
\begin{eqnarray}
\epsilon=abs\big(\frac{\mathcal{U}_{est}-\mathcal{U}}{\mathcal{U}}\big),
\end{eqnarray}
where $abs(\cdot)$ returns the absolute value of a number.  Table \ref{tab_acc} gives the statistic results averaged from 100 tests to demonstrate  how $\gamma$ affects these two indexes. As is shown, with smaller $\gamma$, the estimation error increases and the time cost decreases. To balance the two indexes and provide reasonable estimation accuracy, we set $\gamma=1/4$ in EBUD.

\begin{table}[h] 
	\centering
	\caption{Effect of $\gamma$ on the estimation process\label{tab_acc}} 
	\renewcommand{\arraystretch}{1.2}
	\begin{tabular}{c|c|c|c|c}
		\hline 
		$\gamma$ & 1/2 & 1/4 & 1/8 & 1/16\\ \hline 
	    $\epsilon$ & 10.70\% & 16.54\% & 25.89\% & 30.18\%   \\ \hline
	     $T_{est}$ & 55.37 & 29.06 & 16.8 & 9.6    \\	\hline 
	\end{tabular}
\end{table}


In the deactivation process, each frame only consists of the transmission of reader's request command $Queryd(R_i,~f_i,~PV)$. The time cost is calculated by 
\begin{eqnarray}\label{eq_tdea}
T_{dea}=\sum_{i=1}^{\mathcal{F}_d}\bigg\lceil\frac{f_i+16\times 3+4}{96} \bigg\rceil t_{id}
\end{eqnarray}
Substituting \eqref{eq_test} and \eqref{eq_tdea} into \eqref{eq_T1}, time cost of Phase I is obtained.

\subsection{Time cost of Phase II}\label{sec_timeII}
In Phase II, a new $B$-ary tree-splitting-based missing tag identification (TSMTI) method is developed to quickly identify missing tags. Time cost of Phase II consists of two parts, \textit{i.e.},
\begin{eqnarray}\label{eq_T2}
T_2=T_r+T_t,
\end{eqnarray}
where $T_r$ and $T_t$ are time costs of transmitting reader requests and tag responses in Phase II, respectively.

In frame $F_1$, a tag is randomly assigned into an expected slot indicated in $BV$, and the probability is given by $1/f_1=1/(\beta\mathcal{K})$. In subsequent frames, tags assigned in the same expected collision slot are split into $B$ subgroups. In Fig. \ref{fig_alg_iden}, the splitting process can be viewed as a single search applied to a tree whose root node has $f_1$ children, and all subsequent nodes have $B$  children. Inspired by \cite{Hush98}, we consider these root nodes for the individual tree searches to be at level 0, and the $i$-th level of the tree can be viewed as the $(i+1)$-th frame in Phase II.  In the $i$-th level, the search probes over subintervals of size $B^{i}$. Thus, a tag is assigned to a specific slot of the $i$-th level given by
\begin{eqnarray}\label{eq_p}
	p=\bigg(\frac{1}{f_1}\bigg)B^{-i}=\frac{1}{\beta\mathcal{K}B^{i}}.
\end{eqnarray}

Then, the probability that $j$ out of $\mathcal{K}$ tags fall into a particular slot of level $i$ is 
\begin{eqnarray}\label{eq_pk}
	P(j,\mathcal{K},i)={\mathcal{K}\choose j}p^j(1-p)^{\mathcal{K}-j}.
\end{eqnarray}
Probabilities that a slot is an expected empty, singleton or collision slot are separately given as follows,
\begin{eqnarray}
	P_{empt}&=&P(0,\mathcal{K},i)=(1-p)^\mathcal{K}, \label{eq_pempt} \\
	P_{sing}&=&P(1,\mathcal{K},i)=\mathcal{K}p(1-p)^{\mathcal{K}-1},\label{eq_psing}  \\
	P_{coll}&=&P(j>1,\mathcal{K},i) \label{eq_pcoll}\\ \nonumber
	&=&1-P(0,\mathcal{K},i)-P(1,\mathcal{K},i) \\\nonumber
	&=&1-(1-p)^\mathcal{K}-\mathcal{K}p(1-p)^{\mathcal{K}-1}.
\end{eqnarray}

Let $q_{i}$ be the probability that a particular slot at level $i$ is visited in the splitting process. In level $i$, a slot is visited only when its parent experiences a collision. Otherwise, if its parent slot is empty or singleton, it cannot generate subgroups.  Then, we have
\begin{eqnarray}\label{eq_q}
	q_i=\begin{cases}
	P\left( j>1,\mathcal{K},i-1 \right)&		i\ge 1\\
	1&		i=0\\
	\end{cases}.
\end{eqnarray}
It can be noted that all slots at level 0 will be probed, hence $q_0=1$.   In the $i$-th level, the average number of expected slots to be visited is determined by summing $q_i$ over all subintervals which equal to $\beta\mathcal{K}B^i$, \textit{i.e.},
\begin{eqnarray}\label{eq_sk}
	S_i(\mathcal{K})=\beta \mathcal{K} B^iq_i.
\end{eqnarray}

Reader broadcasts $Querym(~R,~f_1,~BV)$ in the first frame or $Querym(~R,~B,~BV)$ in subsequent frames to tags.  When a tag is assigned to an expected singleton slot, it will reply to the reader.  For each frame, number of segments in $BV$ is obtained by \eqref{eq_sk}.  Since each level refers to one frame and the state of each slot is indicated by at most 2 bits in $BV$, time cost for transmitting reader's request commands can be approximated by
\begin{eqnarray}\label{eq_tr}
	T_r=\sum_{i=1}^{\mathcal{F}_m}T_{r\_f_i}=\sum_{i=0}^{\mathcal{F}_m-1}\bigg\lceil\frac{2S_i(\mathcal{K})+52}{96}\bigg\rceil t_{id} \nonumber\\
	=\sum_{i=0}^{\mathcal{F}_m-1}\bigg\lceil  \frac{2\beta \mathcal{K}B^iq_i+52}{96}\bigg\rceil t_{id},
\end{eqnarray} 
where $T_{r\_f_i}$ is time cost for transmitting reader command in the $i$-th frame, and $\mathcal{F}_m$ is the number of frames needed in Phase II.

If a tag is resolved in a level higher than $i$ in the tree, then it will also be resolved in level $i$\cite{Hush98}. Hence, by counting all singleton slots in level $i$, we are accounting for all singleton slots visited up to and including those at level $i$. Number of identified tags in level $i$ is equal to number of singleton slots in $i$ level minus number of singleton slots in level $i-1$. Then, number of identified tags in level $i$ is calculated as:
\begin{eqnarray}\label{eq_idet}
\mathcal{K}_{i}^{*}=\begin{cases}
\beta \mathcal{K}B^i\left[ P\left( 1,\mathcal{K},i \right) -P\left( 1,\mathcal{K},i-1 \right) \right]&		i\ge 1\\
\beta \mathcal{K}B^iP\left( 1,\mathcal{K},0 \right)&		i=0\\
\end{cases}.
\end{eqnarray}

As is shown in Fig. \ref{fig_alg_iden}, with bit-tracking technology, length of tags' response message in each frame equals the number of expected singleton slots indicated in $BV$.  In the $i$-th frame of Phase II, time cost for transmitting tag responses is given by

\begin{eqnarray}\label{eq_tfi}
	T_{t\_f_i}=\bigg\lceil \frac{\mathcal{K}_{i-1}^{*}}{96} \bigg\rceil t_{id}.
\end{eqnarray}

With \eqref{eq_idet} and \eqref{eq_tfi}, we have 
\begin{small}
	\begin{flalign}\label{eq_tt}
	&T_t=\sum_{i=1}^{\mathcal{F}_m}T_{t\_f_i}=\sum_{i=0}^{\mathcal{F}_m-1}{\lceil \frac{\mathcal{K}_{i}^{*}}{96} \rceil t_{id}}\\
	&=\beta\mathcal{K}\bigg\{P\left(1,\mathcal{K},0\right)\!+\!\sum_{i=1}^{\mathcal{F}_m-1}B^i\left[P\left(1,\mathcal{K},i\right)\!-\!P\left(1,\mathcal{K},i-1\right)\right]\bigg\}\nonumber
	\end{flalign}
\end{small}

\begin{figure}[htp]
	\centering
	\includegraphics[width=0.45\textwidth]{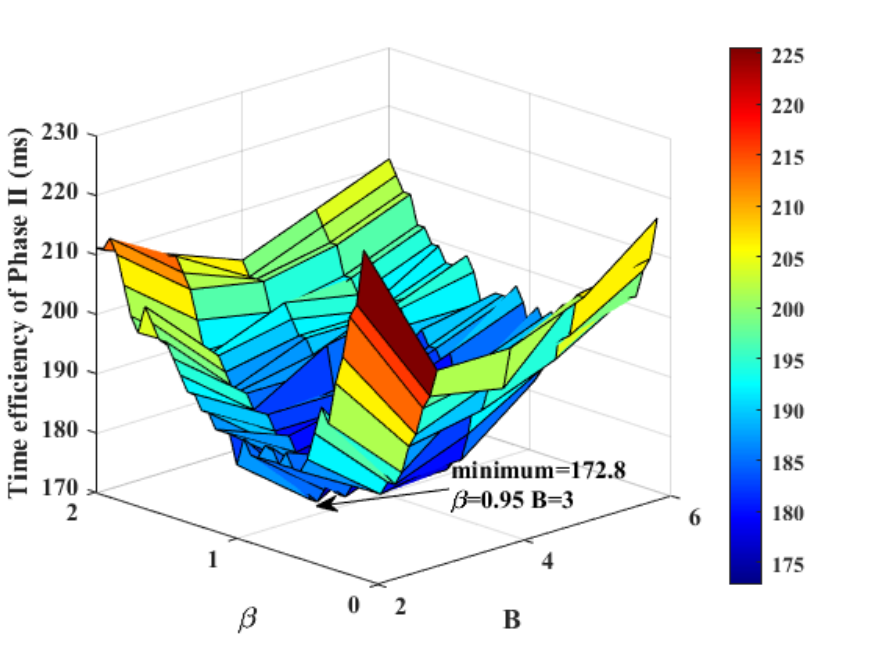}
	\caption{Time cost of Phase II when $\beta$ and $B$ changes.\label{fig_iden_ana}} 					
\end{figure}

Because Phase II terminates when all known tags are identified, $\mathcal{F}_m$ should meet the requirement that
\begin{eqnarray} \label{eq_fm}
 \lceil \sum_{i=0}^{\mathcal{F}_m-1}{\mathcal{K}_{i}^{*}} \rceil =\mathcal{K}\Rightarrow~\mathcal{F}_m. 
\end{eqnarray} 
Substituting \eqref{eq_tr}, \eqref{eq_tt} and \eqref{eq_fm} to \eqref{eq_T2},  time cost of TSMTI is obtained.  In Phase II, two parameters affect the performance of TSMTI, \textit{i.e.}, frame factor $\beta$, and branch number $B$. Fig. \ref{fig_iden_ana} gives the numerical results of $T_{2}$ when $\beta$ and $B$ changes. As can be observed, $T_2$ decreases when $\beta$ ranges from 0.1 to 0.95, and increases when $\beta>0.95$. In the meantime, when $B=3$, $T_2$  is smaller than other settings of $B$. Therefore, the near-optimal parameters are given by $\beta=0.95$ and $B=3$.

\subsection{False negative rate}\label{sec_relia}

In Phase II, if a missing tag is assigned to the expected singleton slot and at least one unknown tags happen to be assigned to the same slot, the missing tag will be falsely identified as present. With \eqref{eq_psing}, number of misidentified missing tags at the $i$-th level of Phase II is given by
\begin{eqnarray}\label{eq_miss}
\mathcal{M}_{fls,i}=\underbrace{\mathcal{M}_i^*}_{missing~tag}\underbrace{ \big[1-(1-p)^{\mathcal{U}_i}\big]}_{unknown~tag}.
\end{eqnarray}
Here, $\mathcal{M}_i^*$ and $\mathcal{U}_i$ are the numbers of missing tags to be identified and unknown tags participating in the $i$-th level of Phase II.  In \eqref{eq_miss}, the first segment represents the number of expected singleton slots with missing tags which equals to $\mathcal{M}_i^*$, and the second segment refers to the probability that at least one unknown tag selects this slot.  Suppose that the missing tags are evenly distributed, the probability that one known tag is missing is $\frac{\mathcal{M}}{\mathcal{K}}$. Based on (\ref{eq_idet}), the number of missing tags to be identified in level $i$ is expressed as:
\begin{eqnarray}\label{eq_idemt}
\mathcal{M}_i^*=\frac{\mathcal{M}}{\mathcal{K}}\cdot\mathcal{K}_{i}^*.
\end{eqnarray}

Only when an unknown tag is assigned to an expected collision slot in level $i-1$, the tag will participate in level $i$. Based on (\ref{eq_pcoll}), the number of remaining unknown tags in level $i$ is given by
\begin{flalign}\label{eq_runk}
	&\mathcal{U}_i=\mathcal{U}_{i-1}P(j>1,\mathcal{K},i-1) \nonumber\\
	&\!\leq\mathcal{U}_0\bigg[1-(1-p)^{\mathcal{K}}\!-\!\mathcal{K}p(1-p)^{\mathcal{K}-1}\bigg].
\end{flalign}
Here $\mathcal{U}_0$ is the number of unknown tags participating in level 0, \textit{i.e.}, the first frame of Phase II.  It equals the number of remaining unknown tags $\mathcal{U}_d$ after Phase I, \textit{i.e.}, $\mathcal{U}_0=\mathcal{U}_d$.
Substituting \eqref{eq_idemt} and \eqref{eq_runk} into \eqref{eq_miss}, number of misidentified missing tags at level $i$ is obtained. Finally, the false negative rate of TSMTI is given by
\begin{flalign}\label{eq_fnr}
&\mathcal{r}_{fn}=\frac{\mathcal{M}_{fls,i}}{\mathcal{M}}=\frac{\sum_{i=0}^{\mathcal{F}_m-1}\mathcal{M}_{fls,i}}{\mathcal{M}}\nonumber \\
&\leq\sum_{i=0}^{\mathcal{F}_m-1}{\frac{\mathcal{K}_{i}^{*}}{\mathcal{K}}}\bigg\{ 1-\left( 1-p \right) ^{\mathcal{U}_d[1-(1-p)^{\mathcal{K}}\!-\!\mathcal{K}p(1-p)^{\mathcal{K}-1}]} \bigg\} .
\end{flalign}

The false negative rate of TSMTI is affected by $\mathcal{U}_d$.  To analyze the effect, we set the remaining unknown tag ratio $r_{ud}=\frac{\mathcal{U}_d}{\mathcal{K}}$, \textit{i.e.}, the percentage of remaining unknown tags to the known ones.  Based on our analysis, Table. \ref{tab_relia_ana} illustrates the numerical values of  $\mathcal{r}_{fn}$ when $r_{ud}$ varies. 
\begin{table}[h] 
	\centering
	\caption{False negative rate of TSMTI when unknown tag ratio varies .\label{tab_relia_ana}} 
	\renewcommand{\arraystretch}{1.2}	
	\resizebox{0.4\textwidth}{!}{\begin{tabular}{c|c|c|c|c|c}
		\hline 
		$r_{ud}$ & 0.01 & 0.05 & 0.10 & 0.15 & 0.20 \\	\hline  $\mathcal{r}_{fn}$ & 0.007 & 0.035 & 0.069 & 0.099 & 0.128 \\ \hline		
	\end{tabular}}
\end{table}

With \eqref{eq_fn}, we have $\mathcal{r}_{fn}<1-\alpha$.  When the required reliability $\alpha=0.9$, $\mathcal{r}_{fn}< 0.1$. According to Table \ref{tab_relia_ana},  the allowed remaining unknown tag ratio $r_{ud}\leq 0.15$ and we set $r_{ud}=0.10$ to meet the requirement.  Similarly, when $\alpha=0.95$ (resp. 0.99), we set $r_{ud}$ is smaller than 0.05 (resp. 0.01), respectively. Therefore, the number of remaining tags should meet

\begin{eqnarray} \label{eq_ud}
\mathcal{U}_d\leq\left\{ \begin{array}{l}
0.1\mathcal{K},~~~ \alpha \leq 0.90\\
0.05\mathcal{K},~~0.90<\alpha \leq 0.95\\
0.01\mathcal{K},~~0.95<\alpha \leq 0.99\\
\end{array} \right. 
\end{eqnarray}

\subsection{Determination of $\mathcal{F}_d$ in Phase I} \label{sec_paraId}

With the required number of remaining unknown tags after Phase I, the number of frames needed to deactivate enough unknown tags can be calculated. Recalling the deactivation process of Phase I,  when an unknown tag is assigned to the expected empty slot indicated in $PV$, it will deactivate itself. Thus, in the $i\text{-th}$ frame of the deactivation process, number of newly deactivated unknown tags $\mathcal{U}_i^*$ is given by
\begin{eqnarray}\label{eq_deun}
\mathcal{U}_i^*=\mathcal{U}_i {f_i\choose 1} \frac{1}{f_i}  \bigg(1-\frac{1}{f_i}\bigg)^\mathcal{K}\approx \mathcal{U}_i  e^{-\mathcal{K}/f_i}=\mathcal{U}_i  ~e^{-1},
\end{eqnarray}
where $\mathcal{U}_i$ is the number of unknown tags participating in the  $i\text{-th}$ frame and the frame size $f_i=\mathcal{K}$. The initial value $\mathcal{U}_1$=$\mathcal{U}$. With recursive resolving, number of remaining unknown tags $\mathcal{U}_{d}$ after $\mathcal{F}_d$ frames  can be calculated as follows,
\begin{eqnarray}\label{eq_ui}
\mathcal{U}_{d}&=&\mathcal{U}_{\mathcal{F}_d}-\mathcal{U}_{\mathcal{F}_d}^*=\mathcal{U}_{\mathcal{F}_d}(1-  ~e^{-1})=\mathcal{U}_{\mathcal{F}_d-1}(1-~e^{1})^2 \nonumber \\
&=&\mathcal{U}_1\big (1-e^{-1}\big )^{\mathcal{F}_d} 
=\mathcal{U}\big (1-e^{-1}\big )^{\mathcal{F}_d}.
\end{eqnarray}

With the estimated unknown tag number, $\mathcal{F}_d$ is obatined by,
\begin{eqnarray}\label{eq_fd}
\mathcal{F}_d \approx\frac{\ln \big(\mathcal{U}_d/\mathcal{U}_{est}\big)}{\ln \big(1-e^{-1}\big)}.
\end{eqnarray}
Substituting \eqref{eq_ud} into \eqref{eq_fd}, the number of frames needed in the deactivation process of Phase I is obtained, \textit{i.e.},
\begin{eqnarray} \label{eq_fdd}
\mathcal{F}_d\geq\left\{ \begin{array}{l}
\frac{\ln \left( \frac{0.1\mathcal{K}}{\mathcal{U}_{est}} \right)}{\ln(1-e^{-1})} ,~~\alpha \leq 0.9\\
\frac{\ln \left( \frac{0.05\mathcal{K}}{\mathcal{U}_{est}} \right)}{\ln(1-e^{-1})} ,~0.9<\alpha \leq 0.95\\
\frac{\ln \left( \frac{0.01\mathcal{K}}{\mathcal{U}_{est}} \right)}{\ln(1-e^{-1})} ,~0.95<\alpha \leq0.99\\
\end{array} \right. 
\end{eqnarray}

%

In conclusion, as is shown in Fig. \ref{fig_paraall}, to determine the number of deactivation frames $\mathcal{F}_d$ in Phase I, the reader first executes estimation process to estimate the number of unknown tags $\mathcal{U}_{est}$ with \eqref{eq_est}.  It then calculates $F_d$ with \eqref{eq_fdd} based on the estimated unknown tag number and the reliability requirement.  

\begin{figure*}[htp] 
	\centering
	\subfloat[\label{fig_t1_n95}]{\includegraphics[width=0.32\textwidth]{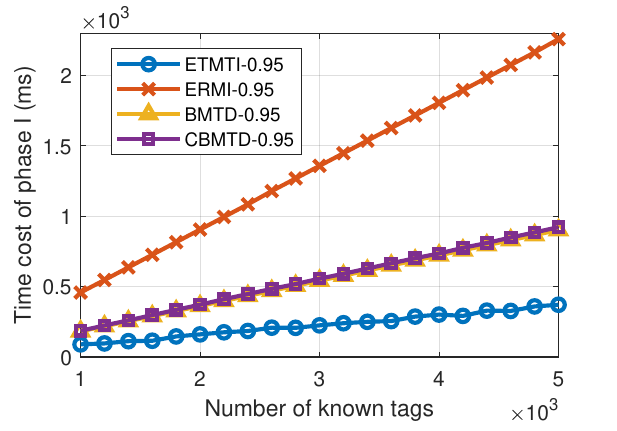}}
	\hfil
	\subfloat[\label{fig_t1_u95}]{\includegraphics[width=0.32\textwidth]{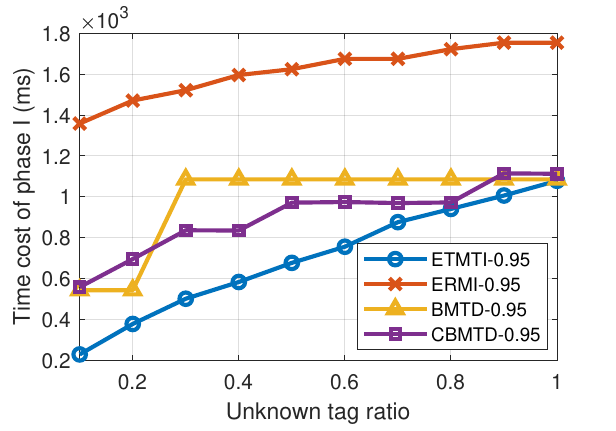}}
	\hfil
	\subfloat[\label{fig_t2_n}]{\includegraphics[width=0.32\textwidth]{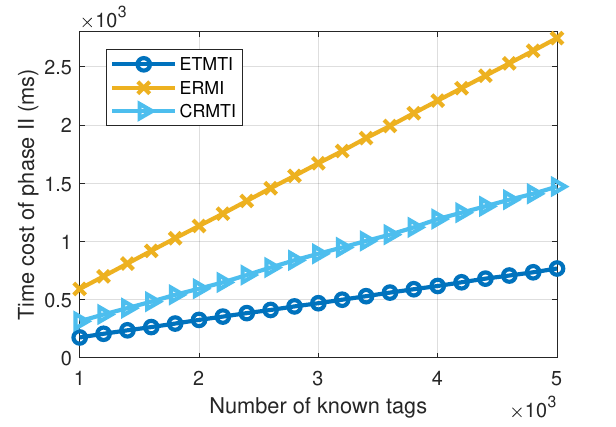}}
	\hfil
	\subfloat[\label{fig_t1_n99}]{\includegraphics[width=0.32\textwidth]{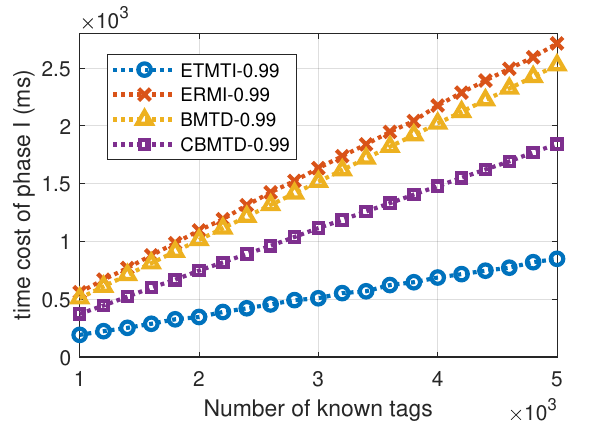}}
	\hfil	
	\subfloat[\label{fig_t1_u99}]{\includegraphics[width=0.32\textwidth]{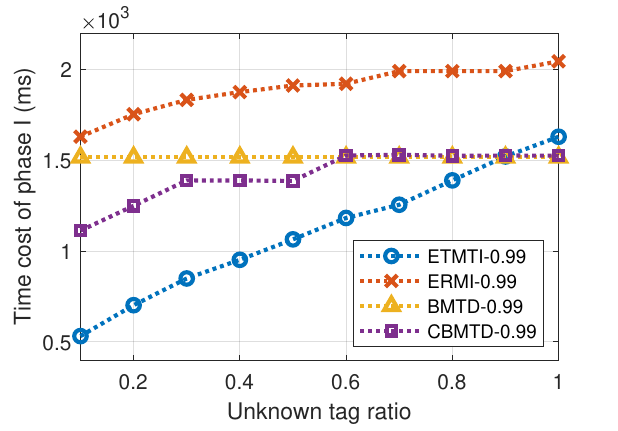}}
	\hfil
	\subfloat[\label{fig_t2_m}]{\includegraphics[width=0.32\textwidth]{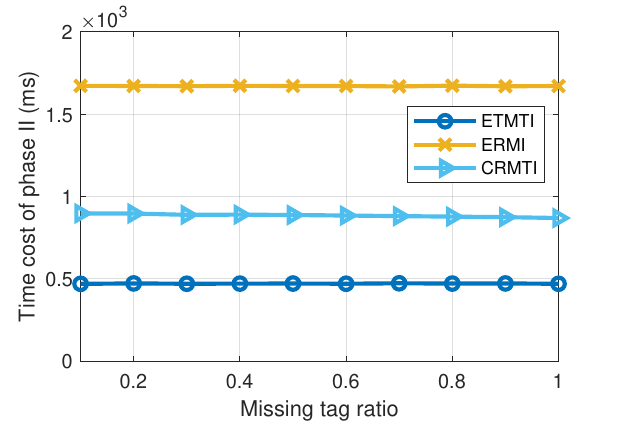}}
	\hfil	
	\caption{Time cost: (a) time cost of Phase I in scenario $S_{11}$; (c) time cost of Phase I in scenario $S_{12}$; (b) time cost of Phase I in scenario $S_{13}$; (e) time cost of Phase I in scenario $S_{14}$; (c) time cost of Phase II in scenario $S_{21}$; (f) time cost of Phase II in scenario $S_{22}$.\label{fig_phaseI}} 				
\end{figure*}

\section{Evaluation} \label{sec_eva}

In this section, we first evaluate the performance of our proposed EIMTI protocol in the deactivation and identification phases, separately. Next, time cost and false negative rate of the overall identification process are given.  Meanwhile,  the results of some best-performing benchmarks are presented for a comprehensive comparison.
\subsection{Simulation configurations}

In the simulation, a typical RFID system that consists of a reader, $\mathcal{K}$ known tags, and $\mathcal{U}$ unknown ones are considered. In a known tag set, $\mathcal{M}$ tags are missing. The reader can retrieve known tags' information from the backend database but has no prior knowledge about the unknown ones. Similar to previous works \cite{Zhang17_PCMTI, Yu17_BMTD,Zhang20_CBMTD}, each tag has a unique ID with 96-bit length, and the data rate between reader and tags is 62.5 Kbps. The transmitted message between reader and tags is divided into a 96-bit segments and each segment takes $t_{id}=2.4$ ms. As in the literature \cite{Chen17_ERMI,Yu17_BMTD,Su23_CRMTI}, communications between reader and tags are assumed to be error-free since the unreliable channel has a similar effect on the comparative benchmarks.

In the simulation, the performance of our proposed ETMTI protocol is compared with the most related ERMI \cite{Chen17_ERMI}, BMTD \cite{Yu17_BMTD}, CBMTD \cite{Zhang20_CBMTD} and CRMTI \cite{Su23_CRMTI}  protocols. ERMI is the most representative missing tag identification protocol that considers the presence of both known and unknown tags. BMTD and CBMTD present the most related unknown tag deactivation methods. CRMTI is the most efficient missing tag identification protocol for situations with only known tags. The simulation is conducted with Matlab R2019b, and each result is averaged over 100 tests.  

\subsection{Time cost of Phase I}

In this phase, the reader estimates the number of unknown tags and deactivates them with multiple frames. To deactivate enough unknown tags, number of frames of this phase is determined with \eqref{eq_fdd}.  In the simulation, we evaluate time cost of Phase I in four scenarios:
\begin{itemize}
	\item~$S_{11}$: $\alpha=0.95$, $r_u=0.1$ and $\mathcal{K}\in[1000,~5000]$;
	\item~$S_{12}$: $\alpha=0.99$, $r_u=0.1$ and $\mathcal{K}\in[1000,~5000]$;
	\item~$S_{13}$: $\alpha=0.95$, $\mathcal{K}=3000$ and  $r_u\in[0.1, 1]$;
	\item~$S_{14}$: $\alpha=0.99$, $\mathcal{K}=3000$ and  $r_u\in[0.1, 1]$.
\end{itemize} 
Since missing tags do not affect the deactivation process, missing tag ratio $r_m$, \textit{i.e.}, the fraction of number of missing tags to that of known tags, is set to be 0. Time cost of ETMTI in Phase I is compared with the most related ERMI, BMTD, and CBMTD protocols, and the comparative results  are presented in Figs. \ref{fig_phaseI}\subref{fig_t1_n95}, \ref{fig_phaseI}\subref{fig_t1_u95}, \ref{fig_phaseI}\subref{fig_t1_n99} and \ref{fig_phaseI}\subref{fig_t1_u99}. It should be noted that the unknown tag number estimation process  in BMTD and CBMTD is neglected in the simulation since the specified unknown tag number estimation method in their works is complicated and time-consuming.

As shown in Figs. \ref{fig_phaseI}\subref{fig_t1_n95} and \ref{fig_phaseI}\subref{fig_t1_n99}, time cost of Phase I increases with the number of known tags. With a fixed unknown tag ratio, number of unknown tags increases with that of known ones. To meet the required reliability, longer frame length and more frames are needed to deactivate enough unknown tags in the deactivation process. Comparing the simulation results in Fig. \ref{fig_phaseI}\subref{fig_t1_n95} with those in Fig. \ref{fig_phaseI}\subref{fig_t1_n99}, we can observe that with higher reliability requirements, time cost of Phase I  also increases. Among the comparative protocols, ETMTI always takes the shortest time to deactivate enough unknown tags, and ERMI takes the longest time. Thanks to the early-breaking and bit-tracking response strategies in ETMTI, time used for unknown tag number estimation is greatly reduced. Thus, it takes much smaller time than other protocols.  Whereas, ERMI takes more time to estimate the number of  unknown tags since it executes the whole estimation frame. Therefore, ERMI takes  more time than ETMTI.  

Taking advantage of multiple hash functions, BMTD uses bloom filters to deactivate unknown tags. In BMTD, the number of frames is determined by minimizing the overall identification time, and the performance of the deactivation phase is not optimized. As demonstrated in Fig. \ref{fig_phaseI}\subref{fig_t1_n95} and \ref{fig_phaseI}\subref{fig_t1_n99}, BMTD takes a longer time than ETMTI, but shorter time than ERMI. Besides, in order to reduce the number of hash functions used in BMTD,  CBMTD proposed a compressed method to reduce time cost of the deactivation process. However, this method may not always work well. In Fig. \ref{fig_phaseI}\subref{fig_t1_n99},  one can observe that time cost of BMTD is  larger than CBMTD when $\alpha=0.99$.   Whereas, as is shown in Fig. \ref{fig_phaseI}\subref{fig_t1_n95},  BMTD and CBMTD take almost the same time when $\alpha=0.95$.

Next, as demonstrated in Fig. \ref{fig_phaseI}\subref{fig_t1_u95} and \ref{fig_phaseI}\subref{fig_t1_u99} that time cost of Phase I increases with an unknown tag ratio. 
One can observe that ETMTI takes the shortest time and ERMI takes the longest time. Thanks to fewer messages are need to estimate the number of unknown tags resulting in less time cost. Moreover, the estimated tag number and number of frames of the deactivation process in ETMTI are appropriately set. 
In ERMI, the frame size of the estimation process is set to be the number of known tags. With a slot-by-slot reply method, more time is needed to estimate unknown tags. Therefore, ERMI takes longer time than ETMTI.
In BMTD, a few frames are used in the deactivation process, but the frame length is set to be very long to deactivate more unknown tags in each frame. Thus, it takes more time than ETMTI, especially when the unknown tag ratio is small.  With compressed filters, CBMTD takes a shorter time than BMTD in most cases. 
In general, the proposed ETMTI protocol shows better performance than other comparative protocols to deactivate unknown tags.

\subsection{Time cost of Phase II}

In this phase, a missing tag identification protocol is executed to verify the presence of known tags and identify missing ones. We evaluate time cost of Phase II in two scenarios: 
\begin{itemize}
	\item~$S_{21}$: $r_m=0.3$ and $\mathcal{K}\in [1000, 5000]$;
	\item~$S_{22}$: $\mathcal{K}=3000$ and $r_m\in [0.1, 1]$.
\end{itemize}
Since the unknown tags do not affect the time cost of Phase II, $r_u$ is set to be 0.
The simulation results of ETMTI are compared with the most related ERMI and CRMTI protocols.

As is illustrated in Fig. \ref{fig_phaseI}\subref{fig_t2_n}, time cost of the missing tag identification protocols increases with the number of known tags. Among the comparative protocols, ETMTI takes the least time to identify all  tags, and ERMI takes the most time. Besides, as is shown in Fig. \ref{fig_phaseI}\subref{fig_t2_m}, time costs of the comparative protocols keep unchanged when the missing tag ratio changes. In this phase, reader has to verify the presence of all known tags and that the identification time is only affected by the number of known tags. With a fixed $\mathcal{K}$, time cost of Phase II keeps unchanged. In the two scenarios, we observe that ETMTI always takes the least time for missing tag identification of Phase II. The main reasons are as follows.

In ETMTI, a new $B$-ary tree-splitting  method is proposed to split colliding tags into smaller groups in a layered structure. The collision probability reduces as the number of layers increases resulting in an increased utilization of the indicative vector.  Whereas, ERMI and CRMTI adopt the Aloha-based method to randomly assign tags repeatedly. In each frame, the collision probability is high. Although CRMTI uses collision resolving method to increase the utilization of indicative vectors, it still takes longer time than ETMTI.
Moreover, tag response strategies used in the comparative protocols are also different. In ERMI, the tag replies with a 1-bit short response in the expected singleton slot. With collision resolving and bit-tracking strategies, CRMTI allows multiple tags to reply with customized responses simultaneously in the expected resolvable collision slot. Thus, time cost for tag response in CRMTI is smaller than that in ERMI. Extending the bit-tracking strategy to all slots, ETMTI further reduces the overhead of each slot and the time cost of ETMTI is smaller than other comparative protocols.

\subsection{Performance of the overall process}

\begin{figure*}[htp]
	\centering
	\subfloat[\label{fig_t_n}]{\includegraphics[width=0.32\textwidth]{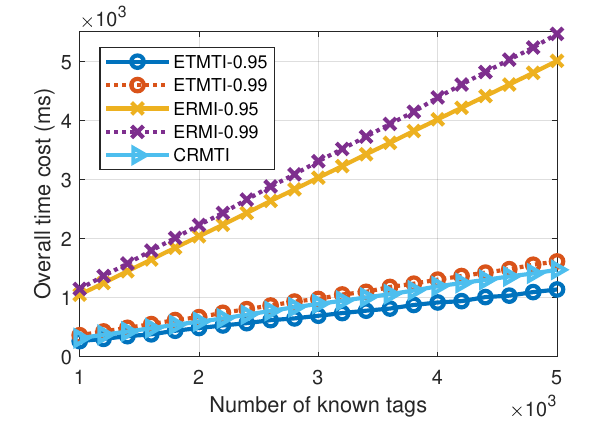}}
	\hfil
	\subfloat[\label{fig_t_m}]{\includegraphics[width=0.32\textwidth]{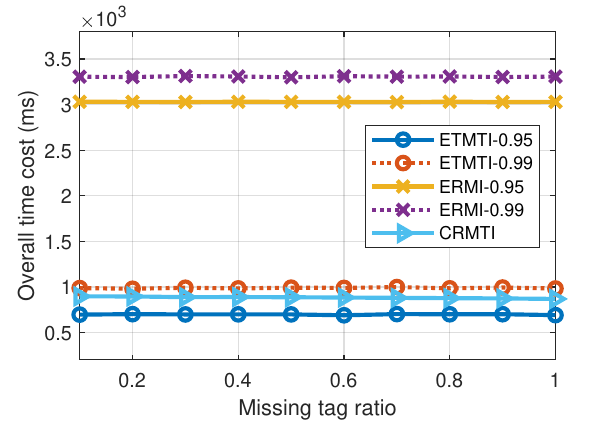}}
	\hfil
	\subfloat[\label{fig_t_u}]{\includegraphics[width=0.32\textwidth]{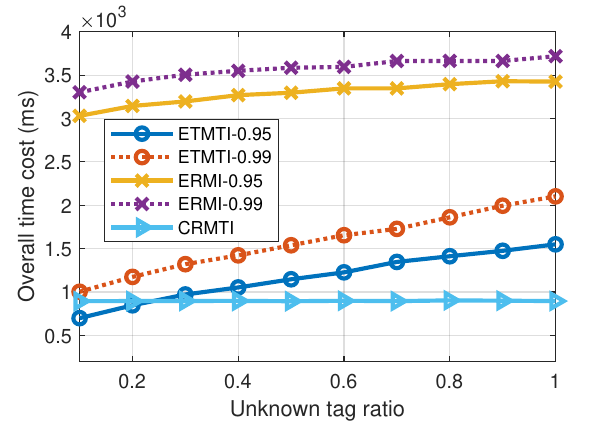}}
	\hfil
	\subfloat[\label{fig_fn_n}]{\includegraphics[width=0.32\textwidth]{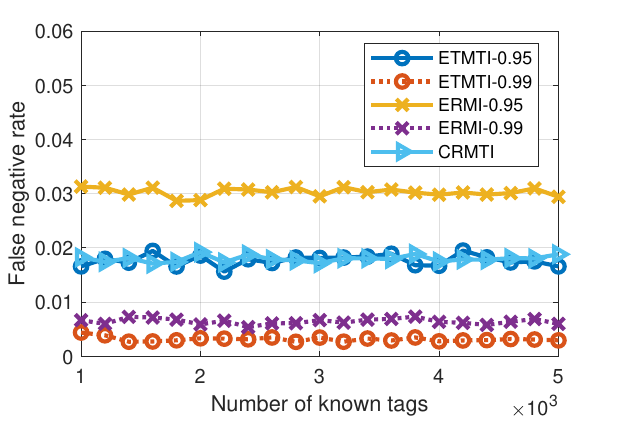}}
	\hfil		
	\subfloat[\label{fig_fn_m}]{\includegraphics[width=0.32\textwidth]{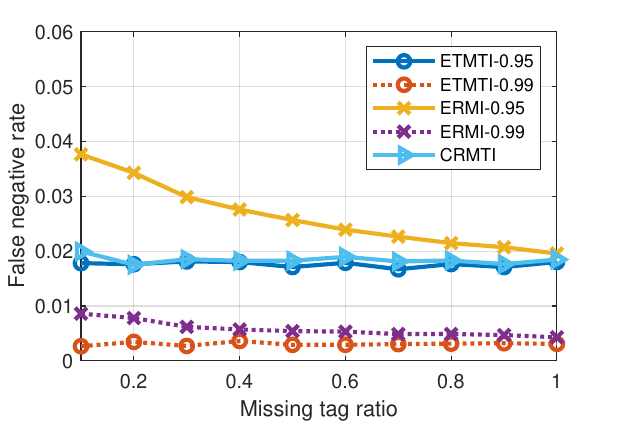}}
	\hfil
	\subfloat[\label{fig_fn_u}]{\includegraphics[width=0.32\textwidth]{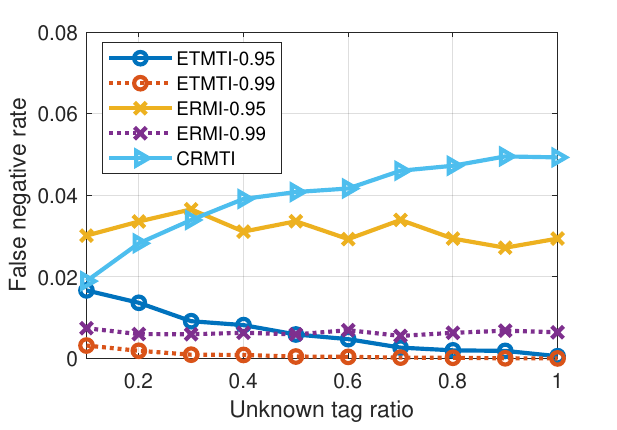}}	
	\caption{Performance of the overall process: (a) time cost vs. number of known tags in scenario $S_{31}$; (b) false negative rate vs. number of known tags in scenario $S_{31}$;  (c) time cost  vs. missing tag ratio in scenario $S_{32}$;  (d) false negative rate vs. missing tag ratio in scenario $S_{32}$;  (e) time cost  vs. unknown tag ratio in scenario $S_{33}$; (d) false negative rate vs. unknown tag ratio in scenario $S_{33}$.  \label{fig_all}} 					
\end{figure*}

In this part, we evaluate the time cost and false negative rate of the overall process in three scenarios:
\begin{itemize}
	\item~$S_{31}$:  $r_m=0.3,~r_u=0.1~\text{and}~\mathcal{K}=[1000,~5000]$;
	\item~$S_{32}$: $\mathcal{K}=3000,~r_u=0.1~\text{and}~r_m=[0.1,~1]$;
	\item~$S_{33}$: $\mathcal{K}=3000,~r_m=0.3,~r_u=[0.1,~1]$;
\end{itemize}
The performance of ETMTI is compared with the most related ERMI and CRMTI protocols and the comparative results are illustrated in Fig. \ref{fig_all}. For ETMTI and ERMI,  simulation experiments when $\alpha=0.95$ and $\alpha=0.99$ are separately conducted in each scenario.

Firstly, as is shown in Fig. \ref{fig_all}\subref{fig_t_n}, the overall time costs of all protocols increase with the number of known tags. Benefiting from the bit-tracking strategies, ETMTI and CRMTI take a much shorter time than ERMI. When $\alpha=0.95$, ETMTI takes the least time than other comparative protocols. When $\alpha=0.99$, ETMTI takes a little bit longer time than CRMTI. As can be observed in Fig. \ref{fig_all}\subref{fig_t_n}, time costs of ETMTI and ERMI increase with the required reliability. With larger $\alpha$, more time is needed in Phase I to deactivate enough unknown tags. 

Fig. \ref{fig_all}\subref{fig_fn_n} presents that false negative rates of all comparative protocols keep unchanged when number of known tags varies. As is shown, false negative rates of ETMTI and ERMI decrease as the required reliability increases. When $\alpha=0.95$, the false negative rate of ERMI is about 0.03, and that of ETMTI is reduced below 0.02. When $\alpha=0.99$, the false negative rate of ERMI is about 0.007, and that of ETMTI is about 0.004. Both ETMTI and ERMI achieve the required reliability, and ETMTI always has a smaller false negative rate than ERMI in the same condition. We can also observe that the false negative rate of CRMTI is almost the same as that of ETMTI (when $\alpha=0.95$). In this scenario, the unknown tag ratio is so small that the unknown tags have little effect on the identification process of CRMTI. Thus, the false negative rate is low. Since CRMTI does not deal with unknown tags, the lowest false negative rate it can achieve is around 0.19.

Secondly, Figs. \ref{fig_all}\subref{fig_t_m} and \ref{fig_all}\subref{fig_fn_m} present the overall time cost and false negative rate separately when the  missing tag ratio varies. As is demonstrated in Fig. \ref{fig_all}\subref{fig_t_m}, time costs of the comparative protocols keep unchanged when the missing tag ratio increases. In ERMI and ETMTI, the overall process is affected by the number of known and unknown tags, as well as the required reliability.  with larger $\alpha$, time costs of ETMTI and ERMI increase since the reader needs more time to deactivate enough unknown tags. CRMTI is only affected by the number of known tags. Therefore, the overall time costs of the comparative protocols do not change with the missing tag ratio. In general, we can observe that ETMTI (when $\alpha=0.95$) takes the shortest time, CRMTI takes longer time than ETMTI (when $\alpha=0.95$), but a little bit lower time than ETMTI (when $\alpha=0.99$), and ERMI always takes the longest time.

In Fig. \ref{fig_all}\subref{fig_fn_m}, ETMTI (when $\alpha=0.99$) has the least false negative rates, and ERMI (when $\alpha=0.95$) has the worst performance. We can also observe that CRMTI shows similar performance with ETMTI (when $\alpha=0.95$), but it takes more time as is shown in Fig. \ref{fig_all}\subref{fig_t_m}. When $\alpha=0.99$, ERMI shows a little bit higher false negative rate than ETMTI, but the increased time cost is too much.
It should be noted that the false negative rate of ERMI decreases as the missing tag ratio increases. In an expected singleton slot, if the assigned known tag is missing and one or more unknown tags are assigned to this slot. The missing tag will be falsely identified as present, resulting in a false negative event. In this scenario, number of unknown tags is a fixed small value. As the missing tag number increases, the percentage of falsely identified missing tags decreases so that the false negative rate decreases accordingly.   

Thirdly, Figs. \ref{fig_all}\subref{fig_t_u} and \ref{fig_all}\subref{fig_fn_u} exhibit the overall time cost and false negative rate when the unknown tag ratio changes, respectively. As is shown in Fig. \ref{fig_all}\subref{fig_t_u}, time cost of CRMTI keeps unchanged since it is only affected by the number of known tags. However, in ETMTI and ERMI, as the unknown tag ratio increases, more time is needed to deactivate enough unknown tags in Phase I. Thus, the overall time costs of ETMTI and ERMI increase with an unknown tag ratio. Similarly, their time costs also increase with the required reliability.
Moreover, as is demonstrated in Fig. \ref{fig_all}\subref{fig_t_u}, when the unknown tag ratio is small, ETMTI (when $\alpha=0.95$) takes the least time. As the unknown tag ratio increases, ETMTI (when $\alpha=0.95$) takes more time than CRMTI. 

In Fig. \ref{fig_all}\subref{fig_fn_u}, ETMTI (when $\alpha=0.99$) has the least false negative rate than other comparative protocols. ERMI (when $\alpha=0.99$) has a higher false negative rate than ETMTI (when $\alpha=0.99$). CRMTI and ERMI (when $\alpha=0.95$) show the worst performance. We can observe that the false negative of ETMTI decreases as the unknown tag ratio increases. With more unknown tags, ETMTI needs more frames to deactivate them in Phase I. This is in accordance with the increasing trends of overall time cost in Fig. \ref{fig_all}\subref{fig_t_u}. The increased number of frames further increases the percentage of deactivated unknown tags resulting in a reduced number of unknown tags that participate in Phase II. Therefore, the false negative rate of ETMTI decreases with the increase of the unknown tag ratio.  

In ERMI, the false negative rate also decreases as the unknown tag ratio increases, but the decrease rate is very small. Since  the number of deactivated unknown tags of ERMI is not as much as that in ETMTI, the decreased false negative rate is not obvious. Note that the fluctuations in ERMI are mainly caused by the inaccurate estimate of unknown tag number. Without any deactivation strategy,  the false negative rate of CRMTI increases with an unknown tag ratio. To sum up, ETMTI exhibits better performance in terms of time cost and false negative rate than the comparative benchmark works.

\section{conclusion} \label{sec_conclusion}

In this work, we proposed an efficient ETMTI protocol to identify missing tags with the presence of unexpected unknown tags in large-scale RFID systems. In ETMTI, two new strategies, \textit{i.e.}, EBUD, and TSMTI are developed to effectively deactivate unknown tags and identify missing tags, respectively. With EBUD, reader can estimate the number of unknown tags within a short time and quickly deactivate enough unknown tags to meet the required reliability. With TSMTI, the colliding tags are more efficiently split into smaller groups which increases the identification efficiency. Moreover, a bit-tracking response strategy is designed to allow the identification of multiple tags in one slot which further reduces time cost. Theoretical analysis is  conducted and the optimal parameters in both EBUD and TSMTI are obtained.  Numerous simulation results are presented to demonstrate the effectiveness of ETMTI. In the future, we will consider to implement collision reconciliation and compression techniques to further reduce time cost of the identification process.

%

\balance

\bibliographystyle{ieeetr}
\bibliography{etmti_ref}

\end{document}